\documentclass[twocolumn,showpacs,preprintnumbers,amsmath,amssymb]{revtex4}
\usepackage{graphicx}
\usepackage{dcolumn}
\usepackage{bm}

\begin{document}

\preprint{Xu et al.}

\title{Resonant reflection at magnetic barriers in quantum wires}

\author{Hengyi Xu}\author{T. Heinzel}
\email{thomas.heinzel@uni-duesseldorf.de} \affiliation{Condensed
Matter Physics Laboratory, Heinrich-Heine-Universit\"at,
Universit\"atsstr.1, 40225 D\"usseldorf, Germany} \author{M.
Evaldsson}\author{S. Ihnatsenka}\author{I. V. Zozoulenko}
\affiliation{Solid State Electronics, Department of Science and
Technology, Link\"{o}ping University, 60174 Norrk\"{o}ping, Sweden}
\date{\today}

\begin{abstract}
The conductance of a quantum wire containing a single magnetic
barrier is studied numerically by means of the recursive Greens
function technique. For sufficiently strong and localized barriers,
Fano - type reflection resonances are observed close to the
pinch-off regime. They are attributed to a magnetoelectric
vortex-type quasibound state inside the magnetic barrier that
interferes with an extended mode outside. We furthermore show that
disorder can substantially modify the residual conductance around
the pinch-off regime.
\end{abstract}

\maketitle

\section{\label{sec:1}INTRODUCTION}

Localized magnetic fields that are oriented perpendicular to a
quantum film \cite{Peeters1993,Matulis1994} or a quantum wire
\cite{Majumdar1996,Guo2000,Papp2001a,Papp2001b,Xu2001,Guo2002,Jiang2002,
Zhai2005,Zhai2006}, which are furthermore strongly localized in
transport (\textit{x}-) direction and homogeneous in the transverse
(\textit{y}-) direction are known as \emph{magnetic barriers} (MBs).
They can be realized experimentally by ferromagnetic films on top of
a two-dimensional
\cite{Monzon1997,Johnson1997,Kubrak2000,Vancura2000,Gallagher2001,
Kubrak2001,Cerchez2007} or quasi one-dimensional electron gas
residing in a semiconductor heterostructure: magnetizing the
ferromagnetic film in \textit{x}- direction results in a magnetic
fringe field with a \textit{z}-component localized at the edge of
the film that extends along the \textit{y}-direction. Transport
experiments on MBs in two-dimensional electron gases show a
pronounced positive magnetoresistance as a function of the MB
amplitude \cite{Kubrak2000,Vancura2000,Gallagher2001,
Kubrak2001,Cerchez2007}, which can be interpreted quantitatively in
a classical picture \cite{Vancura2000,Cerchez2007}, where the MB
acts as a filter with a transmission probability that depends upon
the angle of incidence of the electrons. Moreover, Peeters {\it et
al.} \cite{Peeters1993} studied the energy spectrum and the
transmission properties of MBs with a rectangular profile in a
two-dimensional electron gas by analytical means. This calculation
predicts resonant structures in the low energy region due to the
presence of a virtual level \cite{Matulis1994}.

MBs in quantum wires have been the subject of several theoretical
studies recently, which is driven by their potential ability of
parametric spin filtering \cite{Xu2001,Guo2002,Jiang2002,
Zhai2005,Zhai2006}, provided the effective $g$ - factor is
sufficiently large. MBs are thus not only of fundamental interest,
but also have a distinct potential for application in spintronics.
In these numerical studies, resonant features in the conductance are
frequently found, e.g in Fig. 3 of Ref. \cite{Zhai2005}, Fig. 2 of
Ref. \cite{Zhai2006} or Fig. 5 of Ref. \cite{Governale2000}. The
character of these resonances as well as their origin has not been
studied in detail. This, however, is not only of fundamental
interest, but also a prerequisite for possible applications. For
example, the predicted spin polarizations can reach particularly
high values in the proximity of such resonances \cite{Zhai2006}.

Here, we use the recursive Greens function (RGF) technique to
investigate the structure and the conductance of a single MB that
forms in a quantum wire (QWR) below the edge of a ferromagnetic
film. The barrier shapes are adapted from typical experimental
conditions
\cite{Monzon1997,Johnson1997,Kubrak2000,Vancura2000,Gallagher2001,
Kubrak2001,Cerchez2007}. We find that for smooth barriers with a
large spatial extension, the number of transmitted modes drops
stepwise and without resonances as the barrier amplitude increases
or, correspondingly, the Fermi energy is reduced. As the lowest mode
gets reflected, the conductance of weak barriers approaches zero as
a function of decreasing energy, a situation that we denote as
\emph{magnetic pinch-off}. For sufficiently sharp barriers, however,
pronounced dips in the conductance are found close to the pinch-off
regime. From studies of the local density of states (LDOS) in
combination with the spatially resolved occupation probability
densities and current density distributions, we conclude that the
transmission zeroes originate from the interference of an extended
state with a quasi-bound vortex state that is localized inside the
MB. This leads to resonant reflection
\cite{Porod1992,Porod1993,Shao1994} and can be regarded as a type of
Fano resonance \cite{Fano1961, Satanin2005}. Furthermore, we discuss
the influence of disorder on the reflection resonances.


\section{Model and calculation method}

\begin{figure}[tbp]
\includegraphics[scale=0.5]{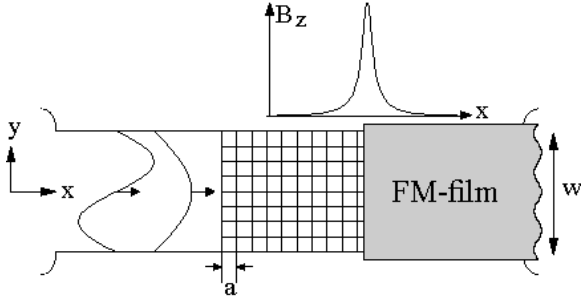}
\caption{Schematic representation of the considered two-terminal
device, consisting of a MB created by a ferromagnetic film deposited
on the top of a quantum wire. Also sketched is the square grid
(period $a = 5\,\mathrm{nm}$) used in the computation.} \label{fig1}
\end{figure}

Let us consider a hard-wall QWR of length $L =4 \,\mathrm{\mu m}$ in \textit{%
x}-direction and width $w =500 \,\mathrm{nm}$ in
\textit{y}-direction, defined in a semiconductor heterostructure
with a ferromagnetic film placed on its surface, Fig. \ref{fig1}. We
use the parameters of GaAs, i.e. an effective electron mass of
$m^*=0.067 m_e$ in our model, where $m_e$ denotes the mass of the
free electron.

The ferromagnetic film has an in-plane magnetization $\mu_0M$ in \textit{x}%
-direction, which generates a symmetric MB $B_z(x)$. In-plane
magnetic
fields are neglected. The inhomogeneous MB can be expressed as \cite%
{Vancura2000}

\begin{figure}[tbp]
\includegraphics[scale=0.5]{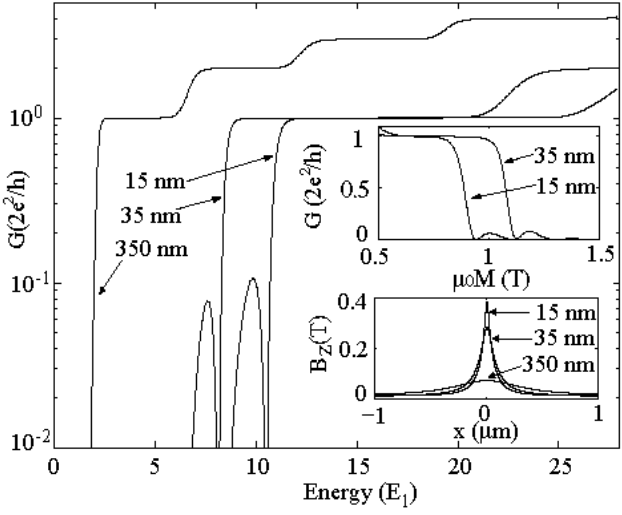}
\caption{The calculated conductance as a function of the Fermi
energy $E_F$ for MBs of different localizations (distances of the
QWR from the sample
surface, calculated for a magnetization of $\protect\mu_0M=1.2\,\mathrm{T}$%
), and as a function of the barrier amplitude for $E_F=
25\,\mathrm{meV}$ (upper inset). The lower inset shows the shapes of
the MBs present at the distances considered, for the film
magnetization assumed in the main figure. Here, $E_1 =
22.5\,\mathrm{\protect\mu eV}$ denotes the ground state energy.}
\label{fig2}
\end{figure}

\begin{equation}
B_z(x) = -\frac{\mu_0 M}{4\pi}\ln\frac{x^2+d^2}{x^2+(d+h)^2}
\end{equation}

\begin{figure}[t]
\includegraphics[scale=0.5]{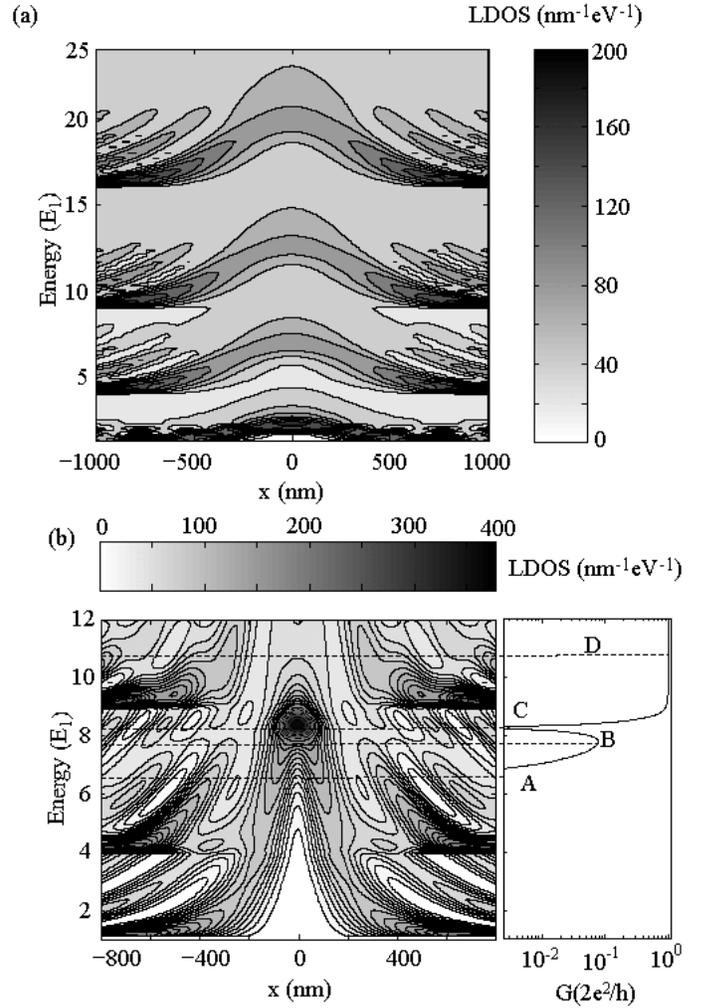}
\caption{The local density of states (LDOS) integrated over the \textit{y}%
-direction, as a function of energy and \textit{x}. (a) For $d=350\,\mathrm{%
nm}$, the modes of the QWR experience a smooth diamagnetic shift in
the barrier region, due to the formation of magnetoelectric subbands
with
spatially varying energy. (b) The LDOS for a strongly localized MB ($d=35\,%
\mathrm{nm}$) shows that the smooth evolution of the suband energies
gets interrupted in the barrier region, and localized states form
around the center of the barrier. Also shown is the position of the
localized state in relation to the conductance of the structure. }
\label{fig3}
\end{figure}

with $h$ being the thickness of ferromagnetic film and $d$ the
distance of the QWR from the semiconductor surface, respectively. As
$d$ increases (which can be realized experimentally by preparing
samples with
two-dimensional electron gases at different locations in the growth (\textit{%
z})-direction), the localization of $B_z(x)$ is reduced. A magnetization of $%
\mu_0M=1.2\,\mathrm{T}$ is assumed, which can be achieved
experimentally by using $Co$ \cite{Vancura2000} or $Dy$
\cite{Cerchez2007} as ferromagnetic material. It can be tuned by
applying an external magnetic field in \textit{x}-direction
\cite{Monzon1997,Johnson1997,Kubrak2000,
Vancura2000,Gallagher2001,Kubrak2001}. The Fermi energy can be
adjusted by, e.g., a homogenous gate electrode in between the
semiconductor surface and
the ferromagnetic film. We consider distances of $d=15\,\mathrm{nm}$, $35\,%
\mathrm{nm}$, and $350\,\mathrm{nm}$, which result in barriers of
amplitudes $B_z(x=0)=0.41\,\mathrm{T}$, $0.28\,\mathrm{T}$,
$0.03\,\mathrm{T}$, and full widths at half maximum (FWHM) of
$94\,\mathrm{nm}$, $148\,\mathrm{nm}$,
and $812\,\mathrm{nm}$, respectively, see the lower inset in Fig. \ref{fig2}%
. Note that the integrated magnetic field $A=\int B_z(x)dx
=6.86\times 10^{-8}\,\mathrm{Tm}$ is independent of $d$.

\begin{figure}[t]
\includegraphics[scale=0.65]{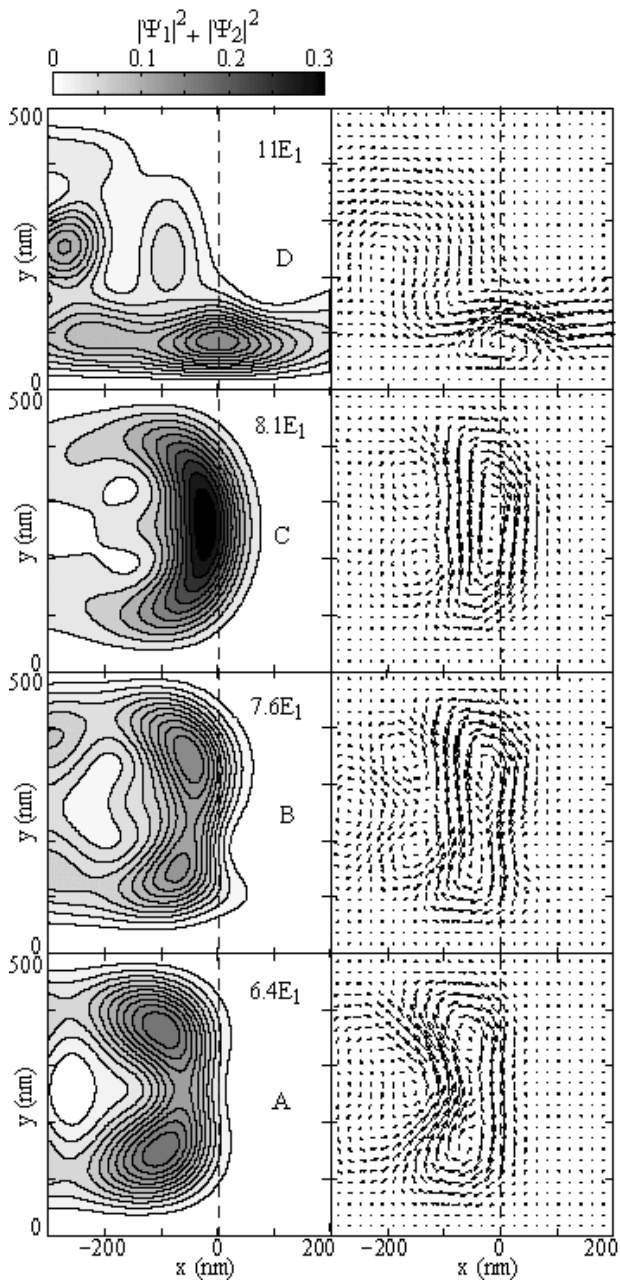}
\caption{Left column: gray scale plots of the sum of the probability
densities $|\Psi_1|^2+|\Psi_2|^2$ of the two occupied modes at the
Fermi energy in the barrier region, shown for the energies marked in
Fig.  \protect \ref{fig2}(b). The maximum of the MB is denoted by
the dashed vertical lines, and its FWHM is $148\,\mathrm{nm}$. Right
column: the corresponding current density distributions. }
\label{fig4}
\end{figure}

The electronic wave functions in the quantum wire exposed to the MB
structure $B_z(x)$ are described by the effective-mass Hamiltonian
\begin{equation}
H = H_0 + V_{c}(y)
\end{equation}
where $V_{c}(y)$ is the confining potential in transverse direction
which is assumed to be a hard-wall potential and $H_0$ is the
kinetic energy term. Using the Landau gauge, the MB can be included
by choosing magnetic vector potential as $\mathbf{A}=(-B_z(x)y, 0,
0)$. The kinetic energy can therefore be written as
\begin{equation}
H_0 = -\frac{\hbar^2}{2m^*}\left[\left(\frac{\partial}{\partial x}-\frac{%
ieB_z(x)y}{\hbar}\right)^2+\frac{\partial^2}{\partial y^2}\right]
\end{equation}

In order to perform numerical computations, the computational area
is discretized into a grid lattice with lattice constant $a =
5\,\mathrm{nm}$ as shown in Fig. \ref{fig1}, such that the
continuous values $x$, $y$ are denoted by discrete variables $ma$,
$na$, respectively. Then the calculation area is connected to two
ideal semi-infinite leads. The tight-binding Hamiltonian of the
system reads
\begin{widetext}
\begin{eqnarray}
H = \sum_m \left\{\sum_n{\epsilon_0}c^\dag_{m,n}c_{m,n}-t\{c^%
\dag_{m,n}c_{m,n+1} +e^{-iqw}c^\dag_{m,n}c_{m+1,n}+\mathrm{H.c.}\}
\right\}
\end{eqnarray}
\end{widetext}
where $\epsilon_0$ is the site energy which has included the effects
of
bottom of the band and confining potential, the hopping element $%
t=\hbar^2/(2m^*a^2)$; $c^\dag_{m,n}$ and $c_{m,n}$ denote the
creation and
annihilation operators at the site $(m,n)$. The phase factor with $q=\frac{e%
}{\hbar}\int ^{x_{i+1}}_{x_i}B_z(x^{\prime}) dx^{\prime}$ is
obtained by using the Peierls substitution.

In the presence of disorder, the site energy changes within a width $\Delta$%
, namely
\begin{equation}
\epsilon_0\rightarrow\epsilon_0+\delta\epsilon_0
\end{equation}
where the values of $\delta\epsilon_0$ are distributed uniformly between $%
-\Delta/2$ and $\Delta/2$ and $\Delta$ is related to the elastic
mean free path $\Lambda$ by \cite{Ando1996}

\begin{equation}
\frac{\Delta }{E_{F}}=(6\lambda _{F}^{3}/\pi ^{3}a^{2}\Lambda
)^{1/2}
\end{equation}%
with $E_{F}$ being the Fermi energy and $\lambda _{F}$ the Fermi
wavelength.

The two-terminal conductance $G_{2t}$  is calculated within the
framework of the Landauer-B\"{u}ttiker formalism and can be
expressed as following \cite{Datta1997}
\begin{equation}
G_{2t}=\frac{2e^{2}}{h}\sum_{\beta ,\alpha =1}^{N}|t_{\beta \alpha
}|^{2}
\end{equation}%
where $N$ is the number of propagating states in the leads,
$t_{\beta \alpha }$ is the transmission amplitude from incoming
state $\alpha $ in the left lead to outgoing state $\beta $. The
transmission amplitudes $t_{\beta \alpha }$ can be expressed via the
total Green's function $\mathbf{G}$ of
the system as $t_{\beta \alpha }=i\hbar \sqrt{v_{\alpha }v_{\beta }}\mathbf{G%
}^{M+1,0},$ where $\mathbf{G}^{M+1,0}$ denotes the matrix $\langle M+1|%
\mathbf{G}|0\rangle $ with $0$ and $M+1$ corresponding to the
positions of the left and right leads. We calculate
$\mathbf{G}^{M+1,0}$on the basis of the recursive Green's function
technique in the hybrid energy space formulation
\cite{Zozoulenko1996a,Zozoulenko1996b}. We calculate separately the
surface Green's functions related to the left and right leads and
the Green's function of the scattering region with the MB, and then
link them together at the boundaries. The wave function $\psi _{i}$
for $i$-th slice of the region under consideration is calculated
recursively via
\begin{equation}
-\psi _{i}=\mathbf{G}^{i0}\mathbf{V}^{10}\psi _{0}+\mathbf{G}^{ii}\mathbf{V}%
^{i,i+1}\psi _{i+1}
\end{equation}%
with $\mathbf{V}$ the hopping matrix and $\mathbf{G}^{i0}$ and $\mathbf{G}%
^{ii}$ the shorthand notations of the Green's functions $\langle i|\mathbf{G}%
|0\rangle $ and $\langle i|\mathbf{G}|i\rangle $. For visualization
purposes, the current density $\mathbf{j_{mn}}$ is associated with
hopping along bonds and can be expressed as
\begin{widetext}
\begin{equation}
j_{mn}(\vec{r})=-i\frac{t}{2a\hbar }\{\mathbf{\widehat{m}}\psi
_{mn}^{\ast }[e^{iqn}\psi _{m+1,n}-e^{iqn}\psi
_{m-1,n}]+\mathbf{\widehat{n}}\psi _{mn}^{\ast }[\psi _{m,n+1}-\psi
_{m,n-1}]-\mathrm{c.c.}\}
\end{equation}
\end{widetext}
where $\psi_{mn}$ is the wave function at site $(m,n)$, and
${\mathbf m},{\mathbf n}$ are the unit vectors in the longitudinal
($x$-) and transverse ($y$-) direction.

The local density of states at the site $\mathbf{r}=(m,n)$ is
related to the total Green's function in real space representation
by the following equation \cite{Datta1997},
\begin{equation}
\rho (\mathbf{r};E)=-\frac{1}{\pi
}\mathrm{Im}[\mathbf{G}(\mathbf{r,r};E)]
\end{equation}%
where $\mathrm{Im}$ means the imaginary part of the complex matrix.

With this formalism, we study below a QWR with a hard wall
confinement potential and a maximum of 4 occupied modes outside the
MB. We set the effective g-factor to zero, therefore the modes
always have a spin degeneracy of 2. Via the distance $d$, the
barrier itself is varied in both amplitude and localization
(characterized by its FWHM) at fixed Fermi energy $E_F$.
Alternatively, we vary $E_F$ and keep the MB constant. In all our
calculations, the electron temperature is set to zero.

\section{\label{sec:3}Results and Interpretation}

In Fig. \ref{fig2}, the numerical results for ballistic QWRs are
summarized. In the upper inset, the conductance of a QWR with 2
occupied modes is shown as a function of the barrier amplitude
$\mu_0 M$, for distances $d$ of $15\,\rm{nm}$ and $35\,\rm{nm}$,
respectively. Quantized conductance steps are observed, which
originate from reflections of the wire modes at the barrier and can
be regarded in close analogy to the conductance quantization in
quantum point contacts \cite{Wees1988,Wharam1988}. Above the
magnetic pinch-off, the conductance shows a pronounced peak with a
maximum conductance up to $\approx 0.1 e^2/h$. We note that the
conductance between the plateau at $2e^2/h$ and the peak equals zero
within numerical accuracy. In the main figure, we show $G$ as a
function of $E_F$ for the barriers given in the inset. For broad
barriers, no structures close to the pinch-off regime are observed.
As the localization of the MB or the number of occupied modes is
increased, the number of conductance zeroes increases as well. Here,
however, we focus on the simplest scenario where just one resonance
is present. This is the reason why we study QWRs with at most four
occupied modes.

To shed some light on the origin of the structures in the
conductance, we study the local density of states (LDOS), integrated
along the \textit{y}-direction, as a function of \textit{x} (Fig.
\ref{fig3}). For smooth MBs (Fig. \ref{fig3}(a)), the MB generates
an \textit{x}-dependent diamagnetic shift of the QWR modes, which
are connected throughout the barrier structure. As the Fermi energy
is lowered, barriers are formed for modes with subsequently lower
energy,
and again a stepwise decrease of $G$ without resonances results.\\
As the barrier is localized further, Fig. \ref{fig3}(b), the modes
segregate and localized states form at the center of the barrier,
which can be regarded as remnants of the magnetoelectric subbands.
Qualitatively, we can understand this as follows. In sufficiently
strong magnetic field gradients, the magnetic phase changes abruptly
in \textit{x}-direction, and pronounced reflections occur which
localize the mode inside the barrier. These localized states may
align in energy with the QWR modes of a higher index and a
\textit{resonant reflection} scenario results. The formation of
localized states at the center of the MB is similar to the case of
the double-barrier resonant tunneling (DBRT) structures, where
transmission resonances are related to the existence of the
quasibound states between two barriers. However, the present
mechanism has a different phenomenology than DBRT: (i) in DBRT, the
resonance transmission probability equals 1 for a symmetric
structure. In our system, the transmission peak is much smaller than
1. (ii) DBRT structures do not possess transmission zeroes. In our
sstem, however, the dip in the conductance between the transmission
peak and the first plateau equals zero within numerical accuracy. As
the strength of the magnetic barrier increases, the height of the
peak increases, but the transmission minimum remains at zero. (iii)
For DBRT, the transmission resonances coincide with the quasi-bound
states in energy and correspond to the poles in the complex-energy
plane \cite{Shao1994}. However, in the present case, as can be
observed in Fig. \ref{fig3} (b) (energy \textit{B}), the maximum in
the LDOS is at a different energy than the conductance peak. Rather,
the peak position is at the low energy tail of the bound state. This
phenomenology is the same as that one found in wave guides with
resonators attached \cite{Porod1992,Porod1993,Shao1994}. Following
Shao's discussion \cite{Shao1994}, the resonator contributes a phase
factor $\lambda$ and the transmission amplitude from left to right
can be expressed as

\begin{equation}
t_{rl}=t_{d,rl}+{t_{rs}t_{sl}}/{(\lambda-r_s)} \label{trl}
\end{equation}

with the transmission amplitudes $t_{sl}$ and $t_{rs}$ for being
scattered into (from the left) and out (to the right) of the
resonator and $r_s$ the factor from each reflection back into the
resonator, and $t_{d,rl}$ denotes the direct transmission path
without a detour into the resonator. According to Eq. \eqref{trl},
the transmission amplitude can vanish if both a direct and an
indirect transmission channel via the resonator are present and
interfere with each other. This behavior corresponds to a Fano
resonance \cite{Fano1961}, which is also found in systems related to
ours \cite{Kunze1995,Satanin2005,Vargiamidis2005}.

In the following, we argue that this scenario does exist in a
magnetic barrier under certain circumstances. The direct
transmission channel is formed by an extended state which resembles
an edge state in the region of high magnetic fields, while the bound
state is a vortex state that forms near the center of the magnetic
barrier. The character of such states is thus markedly different
from snake-orbit states known from magnetic field steps with a
change in polarity \cite{Reijniers2002}.

To further illustrate this effect, we have studied the spatially
resolved probability density and the current density distribution
emerging from the two occupied wave functions at the Fermi level
close to the reflection resonance, see Fig. \ref{fig4}, where the
sum of the probability densities $|\Psi_1|^2+|\Psi_2|^2$ of the two
wave functions (belonging to the first and second energy level of
the quantum wire) as well as the corresponding current density
distributions are plotted as a function of the lateral coordinates
$x$ and $y$ for energies \textit{A}-\textit{D}. In the pinch-off
regime at energies below the transmission resonance (for example at
energy $6.4 \times E_1$, energy \textit{A}), the region of high
probability density inside the barrier is well separated in space
from those in the leads. In addition, the probability of finding the
electrons close to the barrier maximum is small. Correspondingly,
the current density gets reflected at the flank of the barrier. As
the energy is increased into the transmission peak (energy
\textit{B}), the region of high probability density moves along the
\textit{x}-direction into the barrier, and a significant probability
density is found even at the barrier maximum. At the same time, the
probability density remains asymmetric about the center of the QWR
in y-direction. Translated into a current density distribution, this
means that inside the transmission peak, a current path evolves
where the electrons enter the barrier region at the upper edge of
the QWR, and while a large part of the electrons gets rejected, a
significant fraction is transmitted, via the lower half of the QWR,
to the right hand side. At the reflection resonance (energy
\textit{C}), the region of high probability density is pushed even
further into the barrier region, but at the same time develops a
strongly symmetric shape about the center of the QWR cross section.
This means that all the current flowing into the barrier gets
reflected back into the QWR. At higher energies, the open regime is
reached, like at energy \textit{D}. It can be distinguished from the
closed regime by the fact that a high probability density inside the
barrier remains at only one edge of the QWR, which at the same time
extends across the whole barrier structure in x-direction. This
structure provides a strong transmission channel for the electrons,
with the current flowing predominantly at the lower right edge
across the magnetic barrier. This behavior is a consequence of the
formation of a local edge state inside the magnetic barrier.\\

Based on these findings, we interpret the resonances in the
conductance as follows: close to the pinch-off regime, i.e. around
$8 E_1$ for $d=35\,\rm{nm}$ in our model calculations, there is
still a small but non-vanishing direct transmission probability
through the magnetic barrier. This can be inferred from comparing
the width of the transition region between the conductance plateaus
1 and 2, which is roughly 2.5 $E_1$ (see Fig. \ref{fig2}). The
corresponding extended state has the character on an edge state in
the magnetic barrier. According to Eq. \eqref{trl}, it interferes
with the indirect transmission amplitude via the vortex-type bound
state present in the magnetic barrier and generates a reflection
resonance.

\begin{figure}
\includegraphics[scale=0.5]{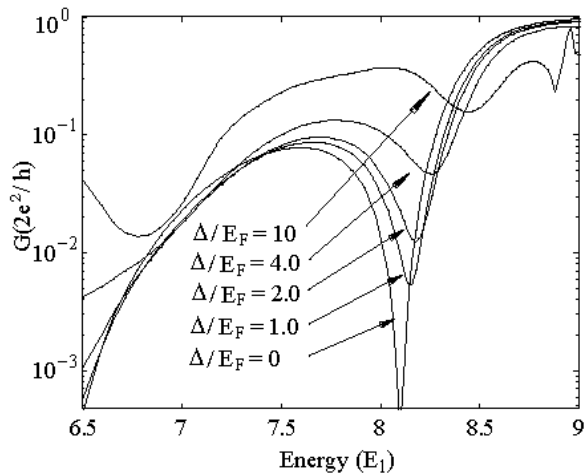}
\caption{Conductance across the magnetic barrier around the
reflection resonance for various degrees of disorder $\Delta/E_F$.
The magnetization of the ferromagnetic film is $\mu_0M
=1.2\,\rm{T}$, and the distance $d$ of the QWR from the surface is
$35\,\rm{nm}$. Note that the parameters without disorder are
identical to those in Fig. \ref{fig2}.} \label{fig5}
\end{figure}

We proceed by discussing the effects of disorder on the transmission
properties. In Fig. \ref{fig5}, the conductance around the
reflection resonance is shown on a logarithmic scale as a function
of the energy for various degrees of disorder. The disorder
potential is modelled by a disorder $\Delta$ of the site energy. A
critical disorder energy of $\Delta\approx E_F$ is observed. For
lower disorder, the reflection resonance remains basically
unaffected. At larger values for the disorder energy, the
transmission at the resonance becomes nonzero, while the minimum in
the conductance shifts towards larger energies. At $\Delta/
E_F\approx 10$, the reflection resonance is no longer a
characteristic property of the structure, while further resonances
are of comparable strength. These resonances have their origin in
interferences due to multiple reflections between impurities. We
emphasize that while the specific disorder configuration at fixed
$\Delta/ E_F$ determines details of the energy-dependent
conductance, it does only marginally influence the position, the
minimum conductance or the width of the reflection resonance.

\section{\label{sec:4}Summary and Conclusion}

The conductance of quantum wires containing a magnetic barrier has
been studied by the recursive Greens function technique. It is found
that for sufficiently large ratios of $\mu_0M / E_F$, the barrier
"closes" and the transmission drop to zero. At energies close to the
magnetic pinch-off, reflection resonances are observed for
sufficiently localized magnetic barriers. The resonances have their
origin in an interference between quasi-bound states residing inside
the magnetic barrier and propagating states of the QWR which, at the
magnetic barrier, have the character of edge states. Even without
taking the spin explicitly into account, it becomes clear that due
to the resonance condition, particularly large spin polarizations
can be expected around the resonances, the sign of which should be
adjustable by a small change of the sample parameters. More studies
are necessary to analyze the details of the spin effects in such
resonances. Furthermore, we hope that our findings stimulate
experimental studies with the objective to observe this type of
reflection resonances.

\begin{acknowledgments}
H.X. and T. H. acknowledge financial support from the Humboldt
Foundation, the Heinrich-Heine Universit\"at D\"usseldorf, and from
the German Academic Exchange Service (DAAD), Aktenzeichen
313-PPP-SE07-lk. S. I. acknowledges support from the Swedish
Institute, and I.Z. acknowledges a DAAD-STINT collaborative grant.
\end{acknowledgments}
\newpage

\end{document}